# Ultracompact Wide-FOV Near-infrared Camera with Wafer-level Manufactured Meta-Aspheric Lens

Chuirong Chi[1], Qichao Hou[1], Guangyuan Zhao[7], Qiang Song[1*,2], Shengyuan Xu[1], Yanling Piao[3], Mengjie Qin[3], Yanan Hu[4], Chaoping Chen[5], Weiwei Cai[6], Yuan Chen[2], Xin Yuan[3,4**], Huigao Duan[1***]

## Abstract

Overcoming the trade-off between wide field of view (FOV) and compactness remains a central challenge for integrating near-infrared (NIR) imaging into smartphones and AR glasses. Existing refractive NIR optics cannot simultaneously achieve ultra-wide angles above 100° and ultrathin total track length (TTL) below 5 mm, limiting their use in portable devices. Here, we present a wafer-level-manufactured meta-aspheric lens (MAL) that achieves a 101.5° FOV, 3.39 mm TTL, and F/1.64 aperture within a compact volume of 0.02 cm³. Unlike previous hybrid lenses with separate refractive and diffractive components, our MAL features a fully integrated structure, which enables a compact form factor. This integration also simplifies fabrication, allowing high-throughput production via micrometer-level precision alignment and bonding on a single wafer, with only one dicing step and no need for additional mechanical fixtures. Furthermore, the design process explicitly considers manufacturability and accurately models metalens dispersion, ensuring that experimental performance matches simulated results. We validate our MAL through both direct and computational imaging experiments. Despite its small form factor, our scalable MAL demonstrates strong NIR imaging performance in blood vessel imaging, eye tracking, and computational pixel super-resolution tasks. This scalable MAL technology establishes a new benchmark for high-performance, miniaturized NIR imaging and opens the door to next-generation smartphone and AR optical systems.

## Introduction

The rapid evolution of smartphone cameras and wearable AR glasses is driving demand for compact, high-performance imaging systems [1–6]. Near-infrared (NIR) imaging, essential for autonomous driving, medical diagnostics, and biometrics, enables deeper tissue penetration and better low-light performance than visible-light modules can deliver[7–11]. However, miniaturizing optical modules for NIR faces major challenges: achieving high imaging quality, ultra-wide field of view (FOV), and extreme compactness simultaneously exceeds the limits of conventional refractive optics. Even state-of-the-art aspheric lens stacks are bulky and costly, and advanced aspheric designs face trade-offs between FOV, aberration correction, and system miniaturization [12–18]. Emerging solutions such as metalenses and planar diffractive optics offer thinner devices and high spatial resolution, but suffer from off-axis aberrations, non-uniform illumination, and limited scalability to wafer-level production [19–23]. Hybrid metasurface-refractive systems have shown some promise; however, current approaches typically employ discrete lens designs that require additional mechanical supports and precise post-fabrication alignment [24–26]. This increases both the lateral dimensions and

1 Greater Bay Area Institute for Innovation, Hunan University, Guangzhou 511300, China
2 Sunny Omni Light Technology Co., Ltd., Shanghai, China
3 Westlake Institute for Optoelectronics, Hangzhou, Zhejiang 311421, China.
4 School of Engineering, Westlake University, Hangzhou, Zhejiang 310030, China
5 State Key Laboratory of Avionics Integration and Aviation System-of-Systems Synthesis, School of Electronic Information and Electrical Engineering, Shanghai Jiao Tong University, Shanghai, China
6 Key Laboratory for Power Machinery and Engineering of the Ministry of Education, School of Mechanical Engineering, Shanghai Jiao Tong University, Shanghai 200240, China
7 Faculty of Engineering, The Chinese University of Hongkong, Hongkong, China.

\* songqiangshanghai@foxmail.com
\*\* xyuan@westlake.edu.cn
\*\*\* duanhg@hnu.edu.cn

assembly complexity, negating the potential for lightweight, compact, and truly mass-producible modules. Furthermore, these discrete hybrid designs are inherently incompatible with standard wafer-level batch fabrication, as accurately assembling discrete diffractive and refractive optics post-dicing is both hard and labor-intensive, restricting their scalability for real-world applications.

Thus, a key bottleneck is the absence of a robust, wafer-level process for monolithically integrating refractive and metasurface elements, which must ensure precise alignment, structural stability, and compatibility with high-throughput manufacturing. Additionally, conventional hybrid diffractive-refractive designs do not account for the true dispersion characteristics of metalenses, resulting in a mismatch between simulated and actual performance, especially for broadband NIR imaging.

Here, we address these challenges with a wafer-level-manufactured meta-aspheric lens (MAL) pipeline that enables monolithic integration of aspheric refractive lenses and metalenses on a single wafer, targeting narrowband NIR imaging with ultra-wide FOV for mobile applications. The MAL is realized by introducing an AA (active-alignment) mask method to bond the aspheric lens and metalens structures at the wafer level, achieving micron-level alignment accuracy and leaving an engineered air gap to maintain the integrity of the nanostructured metasurface. This design eliminates the need for extra mechanical supports and complex post-fabrication alignment, compresses module volume by an order of magnitude, and supports high-yield wafer-scale mass production. Importantly, our optical design process incorporates accurate modeling of metalens dispersion and process integration, capable of precisely calculating the accurate phase response of the metalens under various dimensions such as different wavelengths and incident angles. This makes the model equally applicable to the design of hybrid lenses for broadband near-infrared imaging, enabling close agreement between simulation and fabricated performance, and overcoming the shortcomings of prior hybrid lens design approaches.

This integration achieves a 101.5° FOV, 3.39 mm total track length (TTL), F/1.64 aperture, and a compact volume of 0.02 cm³, offering a substantial total volume reduction over previous hybrid designs while maintaining a modulation transfer function (MTF) exceeding 0.31 at 50 lp/mm, thus reaching the diffraction limit for small-pixel NIR sensors. The MAL achieves uniform relative illumination, over 90% central diffraction efficiency, and close agreement between optical simulations and fabricated results. Unlike previous discrete designs, our approach achieves true monolithic integration, avoiding lateral expansion and ensuring structural stability without auxiliary bonding structures. The MAL balances wide FOV, high MTF, and ultra-compact size more effectively than both pure refractive and metalens-only systems. Meanwhile, compared with the designs of double-layer or multi-layer metalens, the MAL we designed is an improvement on the traditional refractive lens assembly. It can be optimized through geometric optical ray tracing methods, without the need for complex diffraction calculation, and also avoids the reduction in diffraction efficiency and uneven relative illumination caused by the introduction of multi-layer metalens.

We further validate the MAL's performance through direct NIR imaging of blood vessels and eye tracking, as well as computational super-resolution experiments, which integrate the MambaIR model [34], a state-of-the-art neural network for pixel super-resolution and showcases the high compatibility and practical potential of our MAL platform for hardware-software co-optimized NIR imaging. Together, these advances establish a scalable, wafer-level optical solution that bridges high-performance NIR imaging and mass manufacturability, enabling new possibilities for next-generation mobile, AR/VR, and biomedical imaging devices.

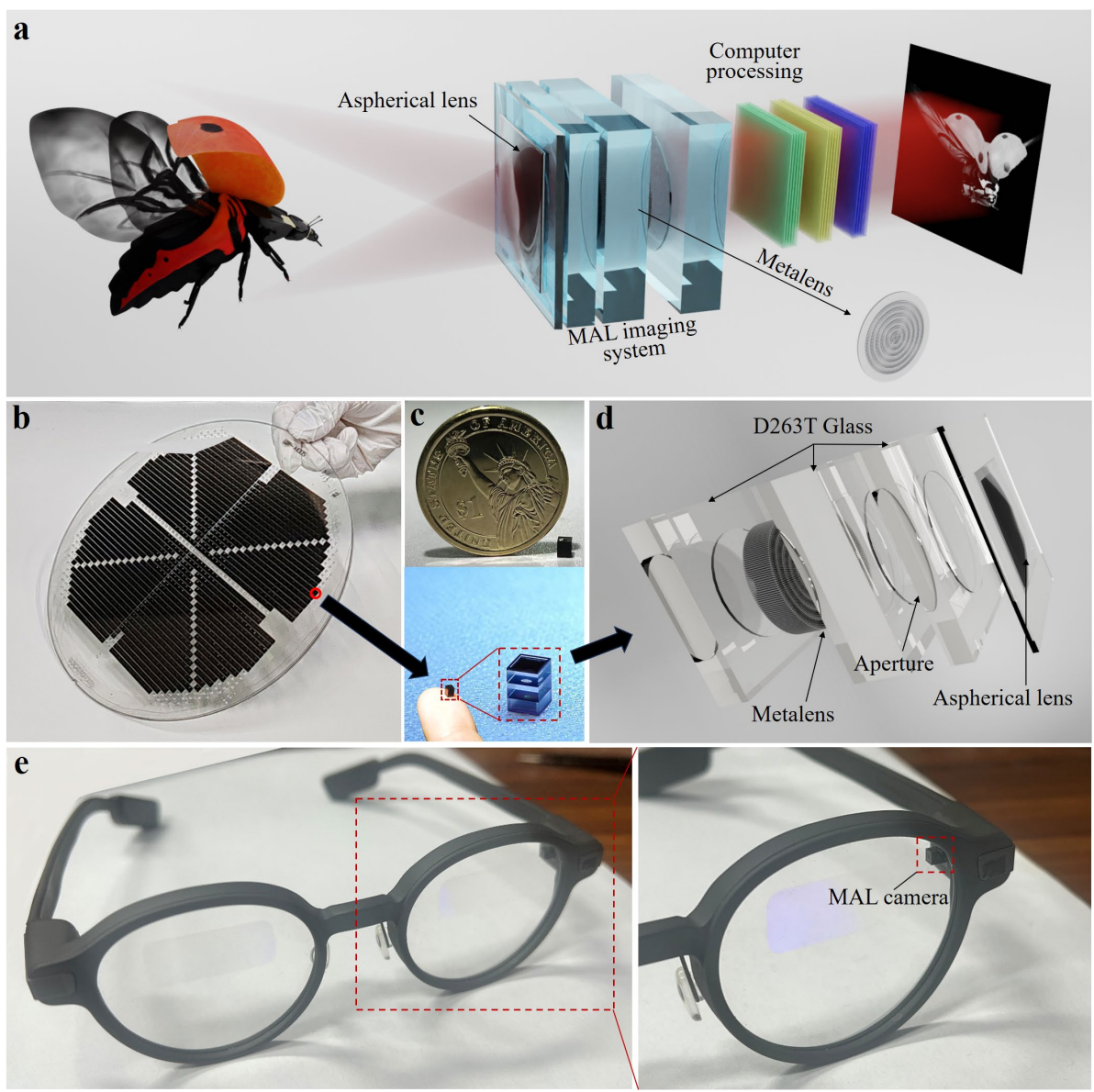

**Fig. 1. Schematic of MAL. a** Illustration of the MAL imaging system. **b** MAL mass-produced on an 8-inch wafer. **c** The size comparison of MAL. **d** The exploded view of MAL. **e** The MAL is integrated into the AR glasses.

## Results
### Schematic of integrated MAL system

The schematic of our integrated imaging system is shown in Fig. 1a. It combines an MAL imaging module and a subsequent image restoration module. The MAL lens is responsible for encoding the target light field onto the sensor, while the image restoration module is designed to refine and enhance the spatial resolution of the acquired images. When tailored to increase the resolution of the image produced by the MAL imaging system automatically, the framework can independently generate a high-fidelity output image with 1K. A comparison of the MAL's size and its exploded view is shown in Fig. 1c and Fig. 1d. It achieves a TTL as low as 3.39 mm and a wide FOV of 101.5° within a minuscule volume (2.564×2.39×3.35 mm$^3$). We compared our MAL with a US one-dollar coin and, which visually demonstrates its MAL's extremely compact size, and thus it has potential to be further integrated into smartphones or wearable AR glasses, as shown in Fig. 1e. Meanwhile, the compact structure of MAL enables batch production of MAL on wafers, facilitating low-cost, rapid manufacturing in large quantities. As shown in Fig. 1b, thousands of MAL can be fabricated at one time on an 8-inch wafer, which greatly promotes the engineering development of high-performance microimaging systems. This advance stems from our integrated new hybrid lens design framework, as shown in Fig. 2a. Unlike conventional MAL systems assembled from discrete elements, our integrated MAL stack attains micrometer-level alignment and intrinsic structural stability without auxiliary mounting hardware, making it fully compatible with high-throughput wafer-level processing. Figure 2b shows the metalens fabricated on the wafer and the bonded MAL structure. The scanning electron microscopy (SEM) images of the fabricated metalens are shown in Fig. 2c. It is notable that the result of manufacturing aligns with the design, as shown in Fig. 2d.

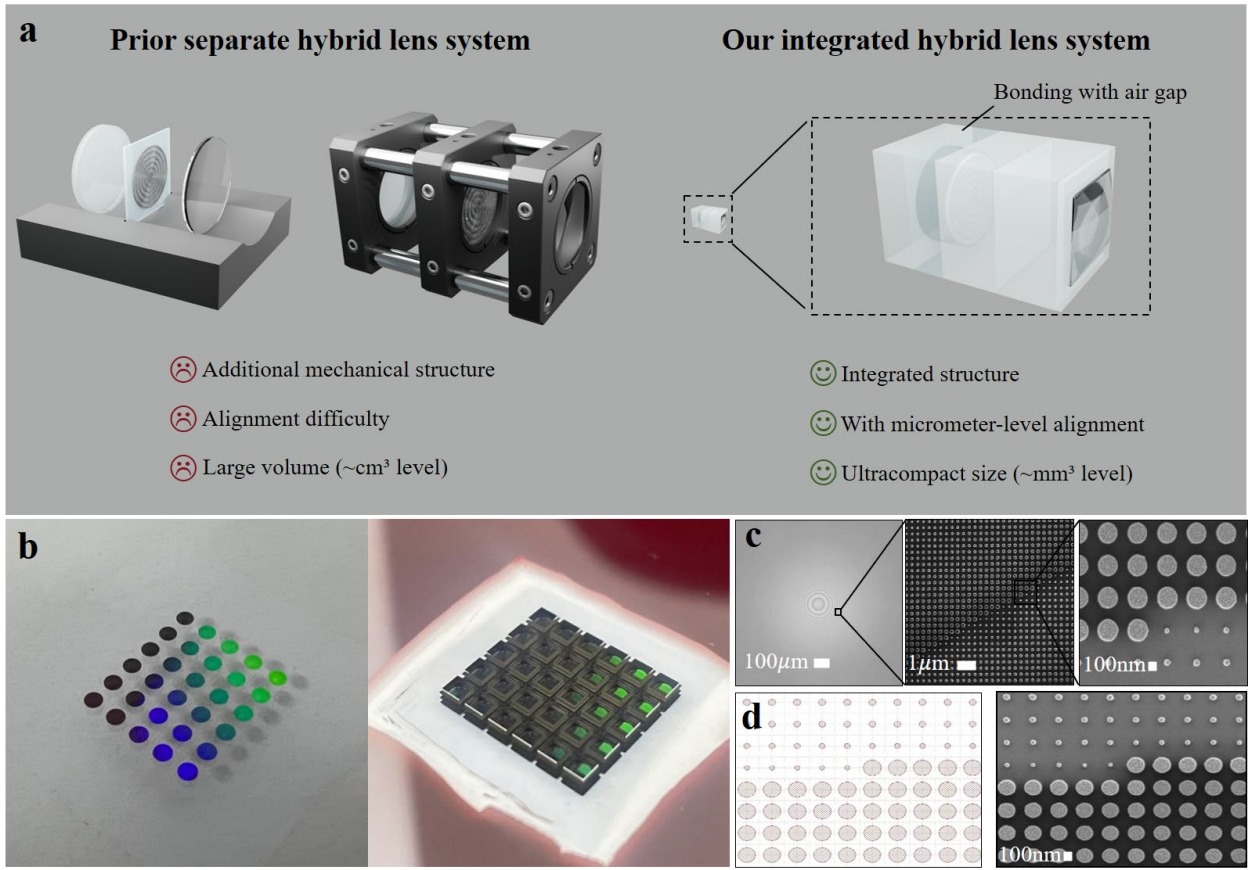

**Fig.2. Schematic diagram and manufacturing process of integrated MAL. a** Comparison of separate and integrated hybrid lens systems. **b** Wafers with metalens arrays and MAL after bonding, taking with mobile photography while the color. **c** The SEM images at different magnifications of metalens. **d** Comparison of metalens layout and fabrication. It is notable that the result of manufacturing aligns with the design.

**Lightfield Analysis**

The performance of MAL was analyzed using Ansys Zemax Optic Studio. To achieve a high-quality image, it is essential to obtain high-contrast and high-resolution images across the entire FOV, as shown in Fig. 3a. Figure. 3b depicts the tangential ray trace of MAL across 0–50.75°. The MTF diagrams of sagittal and tangential beams are presented in Fig. 3c. The MAL demonstrated a high MTF, exceeding 0.5 at 38 lp/mm and 0.31 at 50 lp/mm under the maximum field of view, meeting the diffraction sampling limit for a 640×480 sensor array with 3 μm pixels. We provide MTF diagrams of sagittal and tangential beams at 12 angles within 0−50.75° FOV in note S1 of the Supplementary Materials. These results indicate its capability to provide the necessary high contrast for high-quality imaging. The focusing performance of the MAL was demonstrated through its PSF intensity distribution. As depicted in Fig. 3f, the MAL's PSF displayed a relatively uniform peak intensity in 0°, 9.15°,

27.90°, 43.87°and 50.75°. We showed the PSF intensity distribution at 12 angles within 0−51.51° FOV in note S1 of the Supplementary Materials. Additionally, detailed analyses of distortion, field curvature, spot diagrams, optical path difference, and grid distortion are discussed in note S1 of the Supplementary Materials. It is worth noting that our proposed MAL adopts an integrated design that considers manufacturing process capabilities and structural integration, combining the actual dispersion model of the metalens to calculate the imaging performance of the hybrid lens. Figure. 3d and Fig. 3e show the measured MTF at 50 lp/mm and relative illumination of the MAL, demonstrating that illumination measured in the real experiment and designed through simulation are close, owing to our consideration of the metamaterial dispersion model and the integrated manufacturing design. Additionally, the high-precision Wafer-Level Optics technology is also important in achieving this result. The high degree of aspheric lenses has also achieved high-precision alignment between the measured and designed results, further demonstrating our advantages in design and manufacturing, in contrast to conventional manufacturing methods, where significant discrepancies often occur due to inherent design and processing errors [21]. The relevant data are presented in Supplementary Material S1.

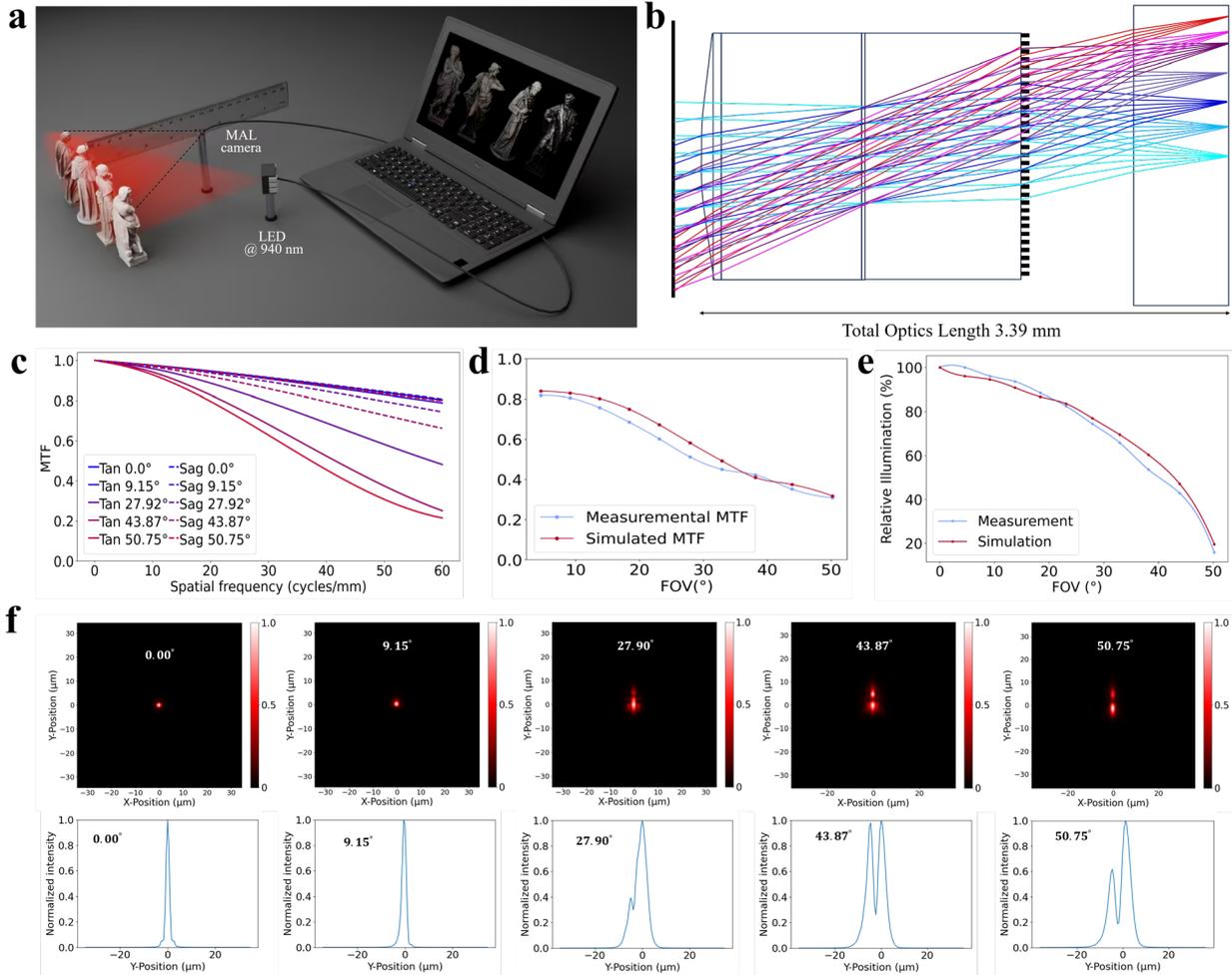

**Fig. 3. The design and result analysis of MAL. a** Schematic diagram of wide FOV MAL imaging system. **b** Ray tracing of different FOVs through the MAL. **c** MTF diagrams include sagittal beams and tangential beams in different FOVs of MAL. **d** The difference between the measuremental and simulated MTF of MAL. **e** The difference between measuremental and simulated relative illumination of MAL. **f** PSF and its Cross-sectional intensity distribution of MAL in five different fields of view. We further show the PSF intensity distribution at 12 angles within $0-50.75°$ FOV in note S1 of the Supplementary Materials.

**Imaging performance analysis**

*Direct Imaging:*

We experimentally test the imaging resolution of the MAL using a USAF resolution test chart. As shown in Fig. 4a, the experimental setup comprises an LED light source, a USAF resolution test chart, and the MAL camera. The image resolution test results are presented in Fig. 4b, where the left panel shows the original USAF image and the right panel displays the corresponding test result. In order to objectively substantiate the achievable spatial resolution, we analyzed each element of the USAF 1951 target in Groups -1 and 0 using only two scalar metrics: Michelson contrast and contrast to noise ratio (CNR). For each element, a centered region of interest (ROI) was extracted and averaged orthogonally to the stripe orientation to obtain a one dimensional intensity profile. Peak and valley intensities were robustly estimated as the mean of local maxima and minima, respectively, while local noise was quantified from the standard deviations within those peak and valley neighborhoods. An element was declared resolvable if Michelson $\geqslant 0.10$ and CNR $\geqslant 2.0$, thresholds consistent with widely used visibility limits for low contrast periodic patterns. Elements with CNR $\geqslant 3$ were additionally labeled as "robust". Under these criteria, the measured ROI in Group -1 exhibited a Michelson contrast of 0.9141 and a CNR of 3.1838, while the ROI in Group 0 exhibited a Michelson contrast of 0.8935 and a CNR of 3.1723. These values are well above the adopted resolvability thresholds, indicating large luminance modulation and low relative noise. Furthermore, we capture images of an eye model from different angles to demonstrate its potential in eye-tracking applications, as illustrated in Fig. 4**c**. The results show that different ocular structures can be clearly distinguished across various imaging angles. Additionally, we evaluated MAL's capability for subcutaneous tissue imaging by comparing hand vascular images acquired independently using a visible light camera and the MAL camera, as shown in Fig. 4**e**. The MAL camera provides clear visualization of dorsal hand veins, which cannot be achieved by conventional visible light imaging. Our MAL demonstrates excellent near-infrared imaging performance. This is because our MAL has a 30nm design bandwidth, which enables excellent imaging results in eye-tracking, vein imaging, and other areas. Moreover, tasks such as face recognition, iris recognition, and fingerprint recognition can also be achieved through MAL, as these applications mainly rely on monochromatic illumination from near-infrared light. Subsequent research will explore the design of larger bandwidths to support applications such as functional imaging, blood oxygen imaging and night vision imaging, which are required for autonomous driving. This will be achieved through the application of our proposed metalens design model featuring practical dispersion. This functionality could be integrated into AR glasses for surgical navigation applications in the future.

*Computational imaging: NIR image super-resolution:*

Our proposed MAL achieves excellent imaging quality and a sufficiently wide field of view (FOV) in the near-infrared band. However, infrared sensors typically have relatively low resolution, resulting in some details being lost in captured images. To address this issue, we introduce computational imaging to further enhance the image quality [27-29]. Computational imaging has emerged as a promising solution for enhancing image quality in applications such as spectral imaging [30-31], image enhancement [32], and super-resolution imaging [33]. Inspired by recent related work, we employed the MambaIR model to enhance the resolution of images captured by the MAL imaging system, thereby enabling more effective information extraction and analysis. It is worth noting that although the MambaIR model has been previously proposed, it was primarily designed for visible light images [34]. To the best of our knowledge, we are the first to adapt and apply it within our system to improve the reconstruction quality of inferred images. Through this approach, we obtained high-quality images with a resolution that was four times higher. As shown in Fig. 4d, compared with the traditional bilinear interpolation method, the MambaIR model can enhance image resolution while reducing the impact of noise, effectively improving the detailed texture and fidelity of the image. To assess the quality of images after super-resolution processing, the BRISQUE score [35], a no-reference image quality assessment method, was employed. This method quantifies image distortion by extracting natural scene statistical features, with scores ranging from 0 to 100, where lower scores indicate better image quality. The details of the MambaIR network model and the evaluation indicators for image quality have been elaborated in detail in the Supplementary Material.

## Discussion

The wafer-level meta-aspheric lens (MAL) marks a significant advance in compact optical system design, resolving long-standing trade-offs among miniaturization, optical performance, and high-volume manufacturability. While previous studies have combined refractive and metasurface optics, they largely treated the metalens as an add-on component to correct aberrations or reduce size within a conventional optical design framework. In contrast, we introduce a co-design strategy that fully integrates the metasurface and aspheric lens from the beginning as a single optical entity. This is enabled by our new concept of "design for hybrid manufacturability," where the optical design proactively incorporates key fabrication constraints—such as bonding interfaces, material dispersion matching, and wafer-scale processes—rather than treating manufacturing as a post-design consideration. As a result, the metasurface and aspheric components are not simply stacked; they are jointly optimized and bonded at the wafer level without needing alignment or mechanical supports. This moves beyond component-level innovation to introduce a system-level methodology that unifies design and mass manufacture, closing the gap between laboratory demonstration and scalable commercial application.

The performance of our MAL is validated across multiple imaging modalities, including direct NIR imaging for blood vessel visualization and eye tracking, as well as computational tasks such as pixel super-resolution and denoising using MambaIR. This comprehensive demonstration confirms the MAL's versatility and effectiveness for both hardware-limited and computation-enhanced imaging scenarios. The synergy between advanced optics and computation is particularly impactful for NIR applications, where both resolution and low noise are critical for uses such as biometrics and medical diagnostics.

Further research will optimize wafer-level fabrication to improve yield and reduce costs, expand integration with advanced computational imaging for challenging conditions, and develop application-specific MAL designs. Moreover, adopting the fully differentiable end-to-end optimization of manufacturability and the downstream optical tasks as well as computational reconstruction method like neural lithography framework could be another exciting next step [36]. Progress in these directions will solidify MAL technology as a cornerstone for the next generation of compact, high-performance imaging systems.

By integrating metalens, aspherical lens, wafer-level manufacturing, and both direct and computational imaging in a single platform, this work redefines the boundaries of compact optical system design. The MAL platform holds the potential to transform mobile devices, AR/VR, medical imaging, navigation, and industrial applications.

## Methods
### Device Design and Manufacturability Co-Optimization

Our meta-aspheric lens (MAL) system is based on a monolithic, wafer-level hybrid architecture that integrates an aspheric lens and a metalens, achieving high optical performance, alignment robustness, and compatibility with scalable mass production. The design was jointly optimized in Zemax

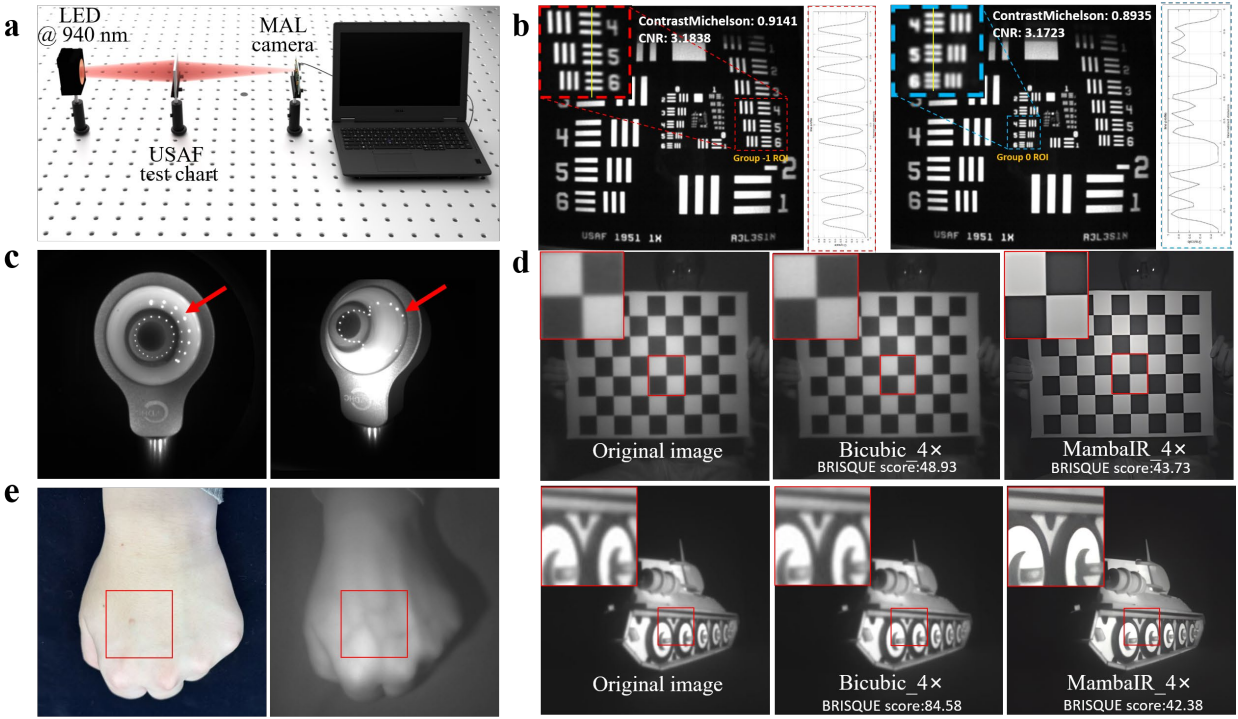

**Fig. 4. Imaging experiments of MAL. a** Experiment setup of USAF 1951 resolution test chart, while **b** imaging results and the quantitative analysis of the different levels of contrast calculation on the resolution plate. **c** Photograph of eye model captured by MAL at different angles. MAL camera. **d** Comparison between the original image, bicubic 4× super-resolution image and the MambaIR 4× super-resolution image of the "Checkerboard" image and the "Tank" image. **e** Comparison of hand images captured by a visible light camera and the MAL camera.

OpticStudio for both imaging performance and manufacturability. System Architecture and Integration Strategy: The MAL employs aspheric and metalens surfaces as the primary functional interfaces, bonded by the D263T glass substrate. The substrate thickness is treated as a design variable during optimization to balance optical performance, compactness, and process constraints. For robust integration, the rear glass surface features an enlarged aperture, and the reserved edge is engineered for structural bonding. A controlled air gap above the metasurface preserves the integrity of the nanostructures during assembly. Aspheric Lens Surface Modeling: The aspheric lens profile is described by:

$$z(r) = \frac{cr^2}{1+\sqrt{1-(1+k)c^2r^2}} + \sum_{n=1}^{m} a_{2n} r^{2n} \quad (1)$$

where $z(r)$ is the surface sag at radial coordinate r, $c=1/R$ is the curvature, k is the conic constant, and $a_{2n}$ are the even-order aspheric coefficients. The surface profile is optimized for minimal aberrations and to ensure process compatibility (see Table S1, Supplementary Materials).

**Metalens Phase and Unit Cell Design:**
The metalens phase distribution is defined by

$$\varphi(x,y) = M \sum_{i=1}^{n} a_n \left(\frac{\rho}{R}\right)^{2i} \quad (2)$$

where $\rho$ is the radial coordinate, $R$ is the normalized metalens radius, and $a_n$ are polynomial coefficients chosen to minimize the focal spot size across the required FOV. The phase profile is empirically controlled with 10 polynomial terms for accurate wavefront shaping. Each phase value is implemented with a circular α-Si nano-pillar meta-atom, simulated via Lumerical FDTD and arranged in a square lattice (400 nm period). The diameter of the nano-pillars ranges from 80 to 290 nm, with a fixed height of 550 nm, enabling continuous phase coverage (0−2π) at high transmittance for the design wavelength of 940 nm (see Fig S2h and Fig S2i, Supplementary Materials). For eachposition (x,y), the optimal unit cell is ass-

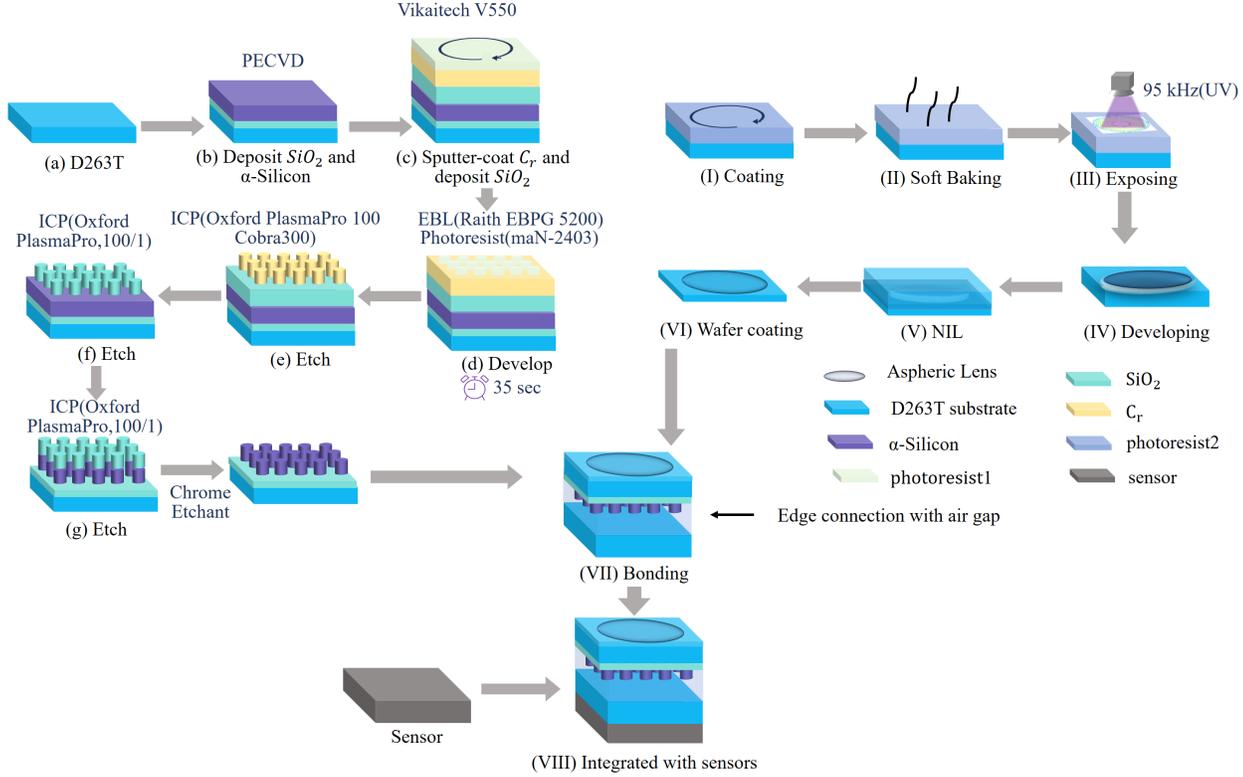

**Fig. 5. Details of MAL fabrication.** The fabrication of MAL includes four main processes: (a)-(g)wafer-level manufacturing of metalens, (I)-(VI)wafer-level manufacturing of aspheric lens, stacking of metalens and aspheric lens, and (VII)-(VIII)integration of the integrated structure with sensors.

-igned from a precomputed database by minimizing a combined phase and transmission error:

$$R(x,y) = \underset{R}{\mathrm{argmin}}(|\varphi_{meta}(x,y,\lambda_0) - \varphi(x,y)| + |T_{meta}(x,y,\lambda_0) - 1|) \quad (3)$$

where $\varphi_{meta}$ and $T_{meta}$ are the phase and transmission for a unit cell at the design wavelength.

**Manufacturability-Aware Dispersion Modeling**:

To ensure simulated performance translates to fabrication, the idealized meta-atom dispersion model is replaced by the experimentally measured dispersion curve of the actual α-Si material. The forward model for local phase modulation then becomes:

$$\varphi(x,y,\lambda,\theta) = f(R(x,y),\lambda,\theta) \quad (4)$$

where $f$ is a bespoke operator that maps the structural distribution, wavelength λ, and incident angle θ to the corresponding phase. This enables rigorous and realistic evaluation of the system's modulation transfer function (MTF), relative illumination, and broadband/off-axis behavior, supporting co-optimization for process compatibility and yield (see Fig S2b and Fig S2c, Supplementary Materials). By incorporating both process-aware geometric constraints and experimentally validated optical models, our design ensures the MAL can deliver high imaging performance and is directly compatible with wafer-scale manufacturing and assembly.

**Device fabrication and integration**

As illustrated in Fig. 5, the fabrication of the meta-aspheric lens (MAL) involves four main stages: (1) wafer-level fabrication of the metalens, (2) wafer-level fabrication of the aspheric lens, (3) stacking and bonding of the two elements with micron-level alignment and engineered air gap, and (4) integration of the MAL structure with the imaging sensor.

1. Metalens Fabrication: The process begins with a 0.5 mm thick D263T Eco glass substrate. A 10 nm SiO$_2$ adhesion layer is first deposited by plasma-enhanced chemical vapor deposition (PECVD), followed by 550 nm of α-silicon. Next, a 30 nm Cr hard mask is sputtered onto the α-silicon surface, and a further 30 nm SiO$_2$ layer is deposited to protect the mask. The designed nanostructure is then patterned onto the wafer using electron beam lithography (EBL, Raith EBPG 5200) with a negative-tone photoresist (maN-2403). After exposure, the resist is developed (ma-D 525, 35s), and the pattern is sequentially transferred: first into the SiO$_2$ (by ICP etching, Oxford PlasmaPro 100 Cobra300), then into the Cr (ICP etching, Oxford PlasmaPro 100/1), and finally into the underlying α-silicon layer (ICP etching, Oxford PlasmaPro 100 Cobra300). The remaining Cr mask is removed by wet chrome etchant, completing the metalens nanostructure.

2. Aspheric Lens Fabrication: In parallel, a non-spherical (aspheric) lens structure is fabricated on a separate wafer using laser direct writing and nanoimprint lithography (NIL). The aspheric lens is formed in photoresist or a UV-curable polymer, as shown in the right branch of Fig. 6: after wafer coating and soft baking, the lens shape is defined by UV exposure (at 95 kHz), followed by resist development and nanoimprint lithography to achieve the final aspheric profile.

3. Stacking and Bonding: After both optical elements are fabricated, wafer-level bonding is performed. The aspheric lens wafer is aligned to the metalens wafer using active mask alignment (AA) with micron-level precision, ensuring accurate registration between the two optical elements. Bonding is accomplished at the wafer edge, deliberately leaving a controlled air gap between the aspheric surface and the underlying metalens to preserve nanostructure integrity and optical performance.

4. Final Integration with Imaging Sensor: The fully integrated MAL structure, now combining the aspheric lens and the metalens, is subsequently diced and mounted onto the imaging sensor to complete the device assembly.

By using this technology, we successfully achieved batch manufacturing of hybrid lenses on wafers. It is worth noting that our process involves presetting mark points on the wafer and using mask alignment for bonding. For each MAL on the wafer, the metalens is primarily connected at the edge and has an air gap in its center. This ensures the stability of the structure and considers the structural stability of the metalens' nanocylinder, while also preventing damage to the original meta cell during the manufacturing process.

To summarize, the above process achieves true wafer-level monolithic integration of aspheric and metasurface optics, enabling high-yield mass production with minimal alignment error and no need for post-fabrication mechanical fixtures. It is worth noting that, we employed EBL for wafer-scale metalens patterning in this work, which is regarded as a technique with sufficient precision but low efficiency, especially in large-scale manufacturing [37]. Our work primarily involves the large-scale manufacturing of hybrid lens systems at the wafer level. For this reason, we have temporarily disregarded the time cost of large-scale manufacturing. Existing processing methods can improve and enhance the production efficiency of metalens at the wafer level [38-41], which can effectively address the challenges in large-scale manufacturing of metalens and achieve scalable manufacturing. In subsequent related work, we will consider improving the manufacturing methods at the wafer level to comprehensively consider multiple factors such as production accuracy, production efficiency, and time cost, in order to achieve more efficient wafer-level manufacturing of micro imaging lenses.

**Image super-resolution reconstruction method**

In this work, we drew on the deep learning methods used for visible light image processing and applied the MambaIR model to our infrared image super-resolution task. As shown in Figure 5a, the low-resolution images were input into the model for convolution to extract shallow features. Subsequently, the feature images passed through modules such as Residual State Space Block (RSSB), Vision State-Space Module (VSSM), and Selective Scan Module (SSM) to achieve deep image feature extraction and fusion, completing high-quality image reconstruction. Leveraging the MambaIR framework, we realized 4× pixel super-resolution, reconstructing raw captures into 1024 × 1024 near-infrared images with high fidelity. Note that the design of MambaIR is almost the same as reference (33) except that visible light images are replaced by NIR images, so details are omitted. If you're interested, you can refer to note S3 in the Supplementary Materials for more details.

**Acknowledgments**
We are grateful to Mr.Junyan Zhu and Mr. Xibian Wei for helpful discussions, and grateful to Xiangbiao Liu from wuhan binary technology Inc. Thanks to Sunny Omni Light Technology Co., Ltd. (Shanghai, China) for financial support.

**Funding**
This work was funded by National Natural Science Foundation of China (grant number 62271414), National Key R&D Program of China(2024YFF0505603), National Key Research and Development Program of China (2022YFB4602600) and National Natural Science Foundation of China (Grant No. 52425508 & 52221001).

**Author contributions**
H. Q. and S. Q. conceived the MAL architecture, conducted theoretical modeling, and led the fabrication of the MAL device. X. S., C. C., G. Z. wrote the manuscript with inputs from the other authors. P.Y. and H.Y. designed the super-resolution algorithm. C. C., C. W., C. Y., Y. X., and D. H. conceived the method and supervised all aspects of the project.

**Data and materials availability**
All data needed to evaluate the conclusions in the paper are present in the paper and/or the Supplementary Materials. Additional data related to this paper may be requested from the authors.

**Declarations**

**Competing interests**
The authors declare that they have no competing interests.